# Magnetic hyperthermia enhances cell toxicity with respect to exogenous heating

*Beatriz Sanz [a,†], M. Pilar Calatayud [a,b], Teobaldo E. Torres [a,b,c], Mónica L. Fanarraga [d], M. Ricardo Ibarra [a,b] and Gerardo F. Goya [a,b,*]*.

[a] Instituto de Nanociencia de Aragón (INA), C/Mariano Esquillor S/N, CP 50018, Universidad de Zaragoza, Zaragoza, Spain.

[b] Departamento de Física de la Materia Condensada, Facultad de Ciencias, C/ Pedro Cerbuna 12, 50009, Zaragoza, Spain.

[c] Laboratorio de Microscopias Avanzadas (LMA), Universidad de Zaragoza, C/Mariano Esquillor S/N, CP 50018, Universidad de Zaragoza, Zaragoza, Spain.

[d] Grupo de Nanomedicina-IDIVAL, Universidad de Cantabria, Herrera Oria s/n, CP 39011 Santander, Spain.

**Corresponding Author**

* E-mail: goya@unizar.es

ABSTRACT: Magnetic hyperthermia is a new type of cancer treatment designed for overcoming resistance to chemotherapy during the treatment of solid, inaccessible human tumors. The main challenge of this technology is increasing the local tumoral temperature with minimal side effects on the surrounding healthy tissue. This work consists of an *in vitro* study that compared the effect of hyperthermia in response to the application of exogenous heating (EHT) sources with the corresponding effect produced by magnetic hyperthermia (MHT) at the same target temperatures. Human neuroblastoma SH-SY5Y cells were loaded with magnetic nanoparticles (MNPs) and packed into dense pellets to generate an environment that is crudely similar to that expected in solid micro-tumors, and the above-mentioned protocols were applied to these cells. These experiments showed that for the same target temperatures, MHT induces a decrease in cell viability that is larger than the corresponding EHT, up to a maximum difference of approximately 45% at T = 46°C. An analysis of the data in terms of temperature efficiency demonstrated that MHT requires an average temperature that is 6°C lower than that required with EHT to produce a similar cytotoxic effect. An analysis of electron microscopy images of the cells after the EHT and MHT treatments indicated that the enhanced effectiveness observed with MHT is associated with local cell destruction triggered by the magnetic nano-heaters. The present study is an essential step toward the development of innovative adjuvant anti-cancer therapies based on local hyperthermia treatments using magnetic particles as nano-heaters.

The use of heat as a therapeutic tool in oncology dates back to the beginning of the previous century.[1] The biological effects of hyperthermia are two-fold: concomitant with the sensitizing effect of heat on tumor cells, which can improve the therapeutic effect of radiation, hyperthermia also produces a direct cytotoxic effect. As a result, an adequate heat treatment can practically kill tumor cells within a nutritionally deficient, hypoxic and acidic environment.[2] Although the therapeutic uses of heat on tumor cells have remained essentially unchanged, the methods for heat generation and delivery have experienced impressive advances. From the first protocols using water baths, heat delivery has evolved to precise, non-contact methods, including radiofrequency, microwaves, and focused ultrasound waves. The use of magnetic nanoparticles (MNPs) as nano-heaters is the most recent of such developments in the delivery of heat to tumor cells and is known as magnetic hyperthermia (MHT) or thermotherapy. The MHT protocol is based on the application of an alternating magnetic field (AMF) to the applied MNPs, which results in the coupling of the magnetic moments of the MNPs with the oscillating field and the conversion of the absorbed energy into heat within the target tumor. The remote, contactless action of the AMF to produce heat locally in the body makes MHT an interesting treatment for eliminating non-accessible, deep tumors that often cannot be treated with surgery. The sublethal heating of cancer cells has been well recognized for many years as a coadjutant therapy to radio- or chemotherapy.[3-5]

The frequencies used for MHT are in the low radiofrequency range, i.e., at the approximately 100-kHz range. The physical mechanism underlying the observed heat generation is the interaction between the magnetic moments of the MNPs and the magnetic component of the applied electromagnetic waves, and the direct interaction of both components with living matter is almost negligible; thus, heat production is only observed in the area where the nano-heaters are delivered in the target region of interest.[6] The rationale behind the use of hyperthermia as a coadjutant protocol for radio- or chemo-therapy is that cancer cells are more sensitive to radiation when they have been previously subjected to high temperatures of 42-45°C, and thus, the use of these two therapies results in a synergistic effect.[2,7]

The physiological effects of temperature on living organisms are well understood.[8] The exposure of mammalian cells to elevated temperatures produces a dose-dependent effect that triggers a cascade of cellular events that compromise and/or damage the cells.[9] In cancer cells, high temperatures also interfere with the regulation of biological processes, such as proliferation and metabolism functions,[10] and the exposure of tissues to temperatures above 43-45°C induces necrosis or apoptosis.[11] Hyperthermia also triggers protein denaturation, which affects the function of most proteins involved in signaling, membrane, cytoskeleton or DNA maintenance, among other functions, and causes cell-growth inhibition and cell apoptosis.[12] Mammalian cells have short-term thermotolerance mechanisms that could be overridden by resistant cancer cells,[13] such as the expression of a heterogeneous family of proteins labeled heat shock proteins (HSPs). These HSPs participate in the process of polypeptide refolding after protein denaturation caused by heat stress.

The neuroblastoma-derived SH-SY5Y cell line is a model of human malignant metastatic neuroblastoma characterized by a high resistance to oxidative[14] and thermal[15] stresses through the overexpression of high levels of HSPs. SH-SY5Y cells therefore exhibit a notable resistance to hyperthermic stress.[16] For this reason, this oncogenic thermo-tolerant cell line was selected as an *in vitro* model for analyzing the effects of magnetic hyperthermia.[17]

This manuscript reports the results of a comparative study of two different heating treatments: heating using an exogenous source, specifically immersion in a water bath (EHT), and local intra-tumoral heating produced by magnetic hyperthermia (MHT). An empirical analysis of the different cellular responses to these two treatments, which focused on the comparison of the cytotoxic effects triggered by MNP-based hyperthermia with the effects induced by exogenous hyperthermia, is then presented.

RESULTS AND DISCUSSION

**Heating treatments: exogenous hyperthermia (EHT) and magnetic hyperthermia (MHT).** This experiment aimed to compare the effects of EHT- and MHT-induced hyperthermia on human neuroblastoma cells. To determine the latent mechanisms of cell death induced by hyperthermia, we produced MHT using in-house synthesized PEI-MNPs as nano-heaters (see Figure 1). Dense cell pellets (see Methods), which roughly simulate a *"micro-tumor"* environment and thus present conditions that resemble some of the intra-tumoral conditions observed during *in vivo* experiments, were used in two comparative analyses, namely immediate ($t_o$) and long-term (t = 6 hours) cell viability assays.

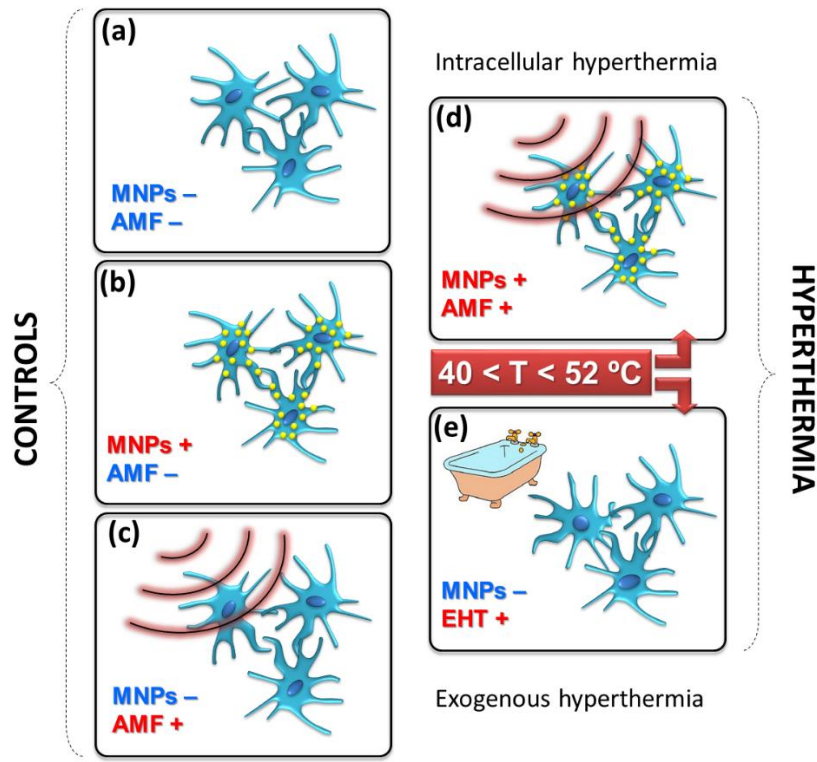

**Figure 1.** Scheme of the thermotherapy assays: the left column represents the three control samples, and the right column presents the experimental treatments with exogenous (EHT) and the local magnetic hyperthermia (AMF).

**Exogenous hyperthermia (EHT).** To investigate the effect of hyperthermia produced by 'exogenous heating' on a set of compacted neuroblastoma cell pellets, these pellets were exogenously heated in a water bath at different target temperatures up to 56°C for 30 minutes and were then allowed to recover in cultures plates immediately after each treatment for 6 hours before their cell viability was measured. The latter analysis aimed to assess any long-term heat-induced cytotoxic effects. These hyperthermia-treated cells were compared with (i) cells treated with AMF, (ii) cells incubated with MNPs but not treated with AMF, and (iii) cells maintained at sub-hyperthermic temperatures (experimental controls).

The control cell samples were maintained at sub-hyperthermic temperatures of 37°C and 40°C and presented viability levels of 97 and 93%, respectively. These results confirmed that sample manipulation induced minor or no effects on cell viability (SI, Figure S1). The viability of cells exposed to mild-temperature hyperthermia (T = 42°C) and then maintained for 6 hours at 37°C in $CO_2$ conditions was also tested, and no changes in cell viability or proliferation were detected. These data allowed us to discard the existence of any 'latent' effect on cell physiology. Incubation of the cells at 43°C resulted in a significant decrease in cell viability to 87% (normalized to the control samples and measured immediately after the hyperthermia treatment), and a viability of 78% was obtained 6 hours after the heating experiment. These divergences between $t_o$ and $t_{6h}$ increased with increasing temperature and reached a maximum difference of 22% with a temperature of 52°C, as shown in the inset of Figure S1 in the SI. For target temperatures above 52°C, a massive cytotoxic effect, with an overall cell viability less than 20%, was observed; thus, no further differences were detected between the early and long-term cytotoxic effects.

In all cases, the viability data as a function of temperature could be fitted by a sigmoidal-type equation:

$$C(T) = \frac{A}{1+e^{B\cdot(T-T_o)}} \qquad \text{Eq. 1}$$

which is a simplified expression based on the two-state model of cell damage developed by Feng et al.[18] for constant exposure times. A more in-depth analysis of Eq. 1 within the context of the two-state model is beyond the scope of the present work, but it is important to note that the phenomenological approach using Eq. 1 allows a straightforward comparison of the parameters that characterize the temperature behaviors obtained with both the EHT and MHT protocols used in this study. The parameter A represents the viability percentage of the control cells (~98%), and B quantifies the temperature width for a given decrease in cell viability. Finally, the parameter $T_0$ (in °C) determines the temperature at which the cell viability function $C(T)$ decreases to 50% of the maximum value. This 50% lethal dose (LD50%) represents the exposure required to kill half of the original cell population. We used Eq. 1 to comparatively analyze the LD50% of both EHT and MHT as a function of increasing target temperatures. The results for EHT showed that it was necessary to reach a temperature of 47.7°C to achieve the long-term effect ($t_{6h}$) of killing 50% of the cells.

**Magnetic hyperthermia (MHT).** Using MNPs as nano-heaters to increase the temperature in 'micro-tumor-phantoms' requires not only detailed knowledge of the power absorption profile of the MNPs but also thermodynamic considerations regarding heat loss in small-scale biological environments.[12,19] To determine the minimum number of cells required for attaining the target temperatures in the hyperthermia region (T < 42°C), we compared several cell pellets containing various numbers of cells (from $2 \times 10^6$ to $2 \times 10^7$ cells in a 100-µL pellet) that were incubated with different concentrations (from 10 to 100 µg/mL) of PEI-MNPs with a SPA=239±19 W/g in water at $f$ = 570 kHz and H = 23.9 kA/m. All the subsequent experiments involved incubation of the cells with 100 µg/mL PEI-MNPs.

To assess the effect of an alternating magnetic field (AMF) on neuroblastoma cells, we exposed the control cells (compacted into dense pellets) to identical experimental conditions ($f$ = 570 kHz; H = 23.9 kA/m; 30 min) in the absence of PEI-MNPs. The cell viabilities at $t_0$ and $t_{6h}$ (Supplementary Information, Figure S2) show negligible cytotoxic effects (97 and 94% viable cells at $t_0$ and $t_{6h}$, respectively). The effects on the cell viability of the magnetically loaded cells were investigated by increasing the T from a starting permissive set point of 37°C. At sub-hyperthermic temperatures of 37 and 40°C, the required amplitude of the AMF was quite small. Accordingly, the two corresponding points in the curve showed that the nano-heaters induced little or no effect, resulting in high viability percentages (91% and 87% at $t_0$, respectively). However, although the differences in cytotoxicity at $t_0$ and $t_{6h}$ were negligible at 37°C, the viability of the cells exposed to T = 40°C decreased from 87% (at $t_0$) to 75% at $t_{6h}$. Increased differences between $t_0$ and $t_{6h}$ were obtained at higher target temperatures, and a maximum difference in viability of 20-24% was found at temperatures near 46°C (SI, Figure S2 inset). Based on the observed differences in cell viability between $t_0$ and $t_{6h}$, we decided to focus on the effects on cell viability at $t_{6h}$ while bearing in mind that the cytotoxic effect of MHT could be significantly enhanced depending on the post-treatment time scale. According to Eq. 1, a $T_0$ value equal to 42.1°C is required for a long-term effect.

Finally, a comparison between the cell viability curves obtained at $t_{6h}$ after the exogenous and localized treatments showed a clear difference in the $T_0$ required to obtain a similar cytotoxic effect on neuroblastoma cells. Figure 2 presents a direct comparison between both tendency curves and demonstrates how the $L_{50}$ of the dense cell pellets treated with MHT was substantially lower (i.e., 42.1°C) than that obtained with exogenous hyperthermia. At this temperature, 88.8% of the neuroblastoma cells in the EHT experiments were still unaffected. Under our experimental conditions, a maximum difference in cytotoxicity of approximately 45% was observed at $T_0 = 46$°C.

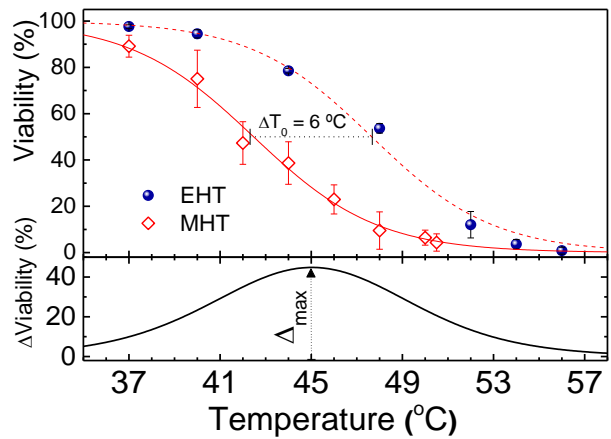

**Figure 2.** Viability after EHT (solid circles) and MHT (open diamonds) treatments (570 kHz; 30 min, 100 µg/mL, n = 5) as a function of the target temperature. The dashed lines are the trend lines predicted using Eq. 1. The difference in the temperatures ($\Delta T_0 = 6$°C) needed to induce 50% cell death is indicated by the dotted line between the EHT and MHT curves. Lower panel: curve of the difference between the EHT and MHT viability data, showing that the maximum viability difference $\Delta_{max} = 45$% was obtained at T = 45.1°C.

**Qualitative and quantitative evaluation of hyperthermia-induced cell death.** In response to an external insult, cells die by apoptosis and/or necrosis.[20] To evaluate the cytotoxic effects of exposure to the heating treatments for 30 minutes in detail, we analyzed the cells by flow cytometry (FACS) using a series of cell indicators of death by necrosis and/or apoptosis (see the materials section). In brief, early phases of apoptosis were identified by the presence of Annexin V on the surface of cells, whereas late apoptotic cells were identified through simultaneous labeling with Annexin V and non-vital dyes. Additionally, necrotic cells were only stained with non-vital dyes. Hence, this study allowed us to quantitatively and qualitatively determine the type of cell death experienced by these neuroblastoma cells. Figure 3 shows the overall percentages of cell death (including apoptotic and necrotic cells) observed in hyperthermia-treated cells exposed to temperature $T_0$ corresponding to the LD50% identified for the EHT protocol. These data have been adapted from the dot plots (see Figure S3 in the supplementary Information) that make possible to differentiate between those cells that express each one of the fluorescent markers from those that express neither or both. As expected, the MHT treatment was found to exert its cytotoxic effect more efficiently than the exogenous hyperthermia treatment (approximately 77% versus 57%). Interestingly, this study also revealed that the two hyperthermia treatments induced different cell death mechanisms (late apoptosis-necrosis).

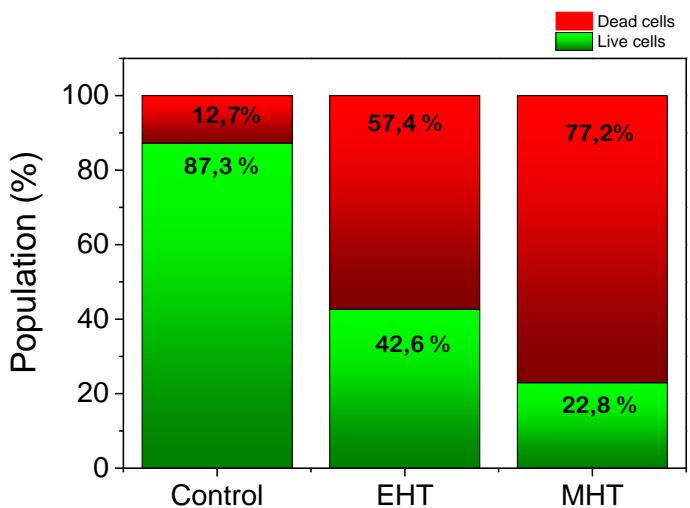

**Figure 3.** Quantitative and qualitative flow cytometric analysis of cell death in control and hyperthermia-treated SH-SY5Y cells incubated with PEI-MNPs (100 μg/mL) at $T_0 = 42.5°C$ for 30 minutes. The viability data are displayed after grouping apoptotic and necrotic death (red

bars) and show that exogenous hyperthermia (EHT) is less effective than magnetic hyperthermia (MHT).

An analysis of confocal microscopy projection images of the nuclei of control and hyperthermia-treated SH-SY5Y cells (see Figure 4) revealed that the untreated control cells display a well-shaped nuclei (A, blue channel) comparable to that of cells exposed to Alexa 488-labeled MNPs (B, red channel). Occasional mitotic figures were also observed (empty arrows). However, irregularly shaped nuclei and occasional compacted chromatin spots indicative of pyknosis were observed in the cells exposed to MNPs and EHT (C). In addition, the cells exposed to magnetic hyperthermia (D) display aberrant nuclear shapes, and some of the cells exhibit some morphological features indicative of karyolysis, pyknosis (blue arrow) and karyorrhexis (red arrow).

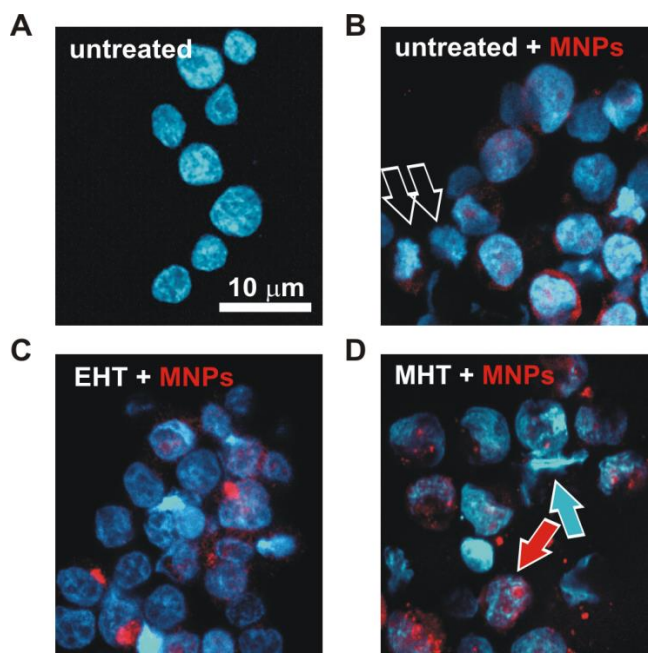

**Figure 4.** Confocal microscopy projection images of SH-SY5Y cells exposed to Alexa 488-labeled MNPs. The nuclei (blue channel) of the untreated cells display a fairly regular circular shape that does not differ from that of the nuclei of cells exposed to f-PEI-MNPs (100 µg/mL,

red channel). In contrast, neuroblastoma cells exposed to hyperthermia show evident nuclear morphological shape aberrations, and these aberrations are more pronounced in the cells treated with MHT (blue arrow). All images are pseudocolored.

The effects of temperature, primarily exogenous hyperthermia treatments, on cell viability have been previously reported by several researchers. Cheng *et al.* studied the effect of hyperthermia by heating SH-SY5Y cells at 43°C for 30 minutes with a water bath and subsequently culturing these cells for 6 hours, similar to the protocol used in the present study.[16] These researchers found that heating at 46.5°C resulted in 50% viability, which is in good agreement with the results of our EHT experiments, and suggested a relationship between thermotolerance and the high expression levels of some heat shock proteins in SH-SY5Y neuroblastoma cells.[16]

**Cellular damage induced by thermotherapy.**

*Exogenous hyperthermia (EHT).* Figure 5 presents TEM and dual-beam (FIB-SEM) images of neuroblastoma cells exposed to EHT hyperthermia treatment at temperatures of 46 (Figure 5B) and 52°C (Figure 5C) for 30 minutes. Compared with the untreated control cells, the cells exposed to hyperthermia typically exhibited complete vacuolization of all organelles in the cytosol, including mitochondria and the endoplasmic reticulum, and a marked rupture of the plasma membrane (Figure 5A). These cell features indicative of necrosis were more evident in the cells exposed to 52°C (Figure 5C), and the cells subjected to EHT presented a clear change in their external morphology, which presented a rough surface and a collapsed appearance.

Using a dual-beam (FIB-SEM) microscope, a series of cross-sectional images (Figures 5D-5F) of cells subjected to EHT at 52°C for 30 minutes showed a "deflated balloon" morphology and a holey membrane. Consistent with the previous discussion, evident cytoplasmic vacuolation and irregularly shaped nuclei were observed in all analyzed samples.

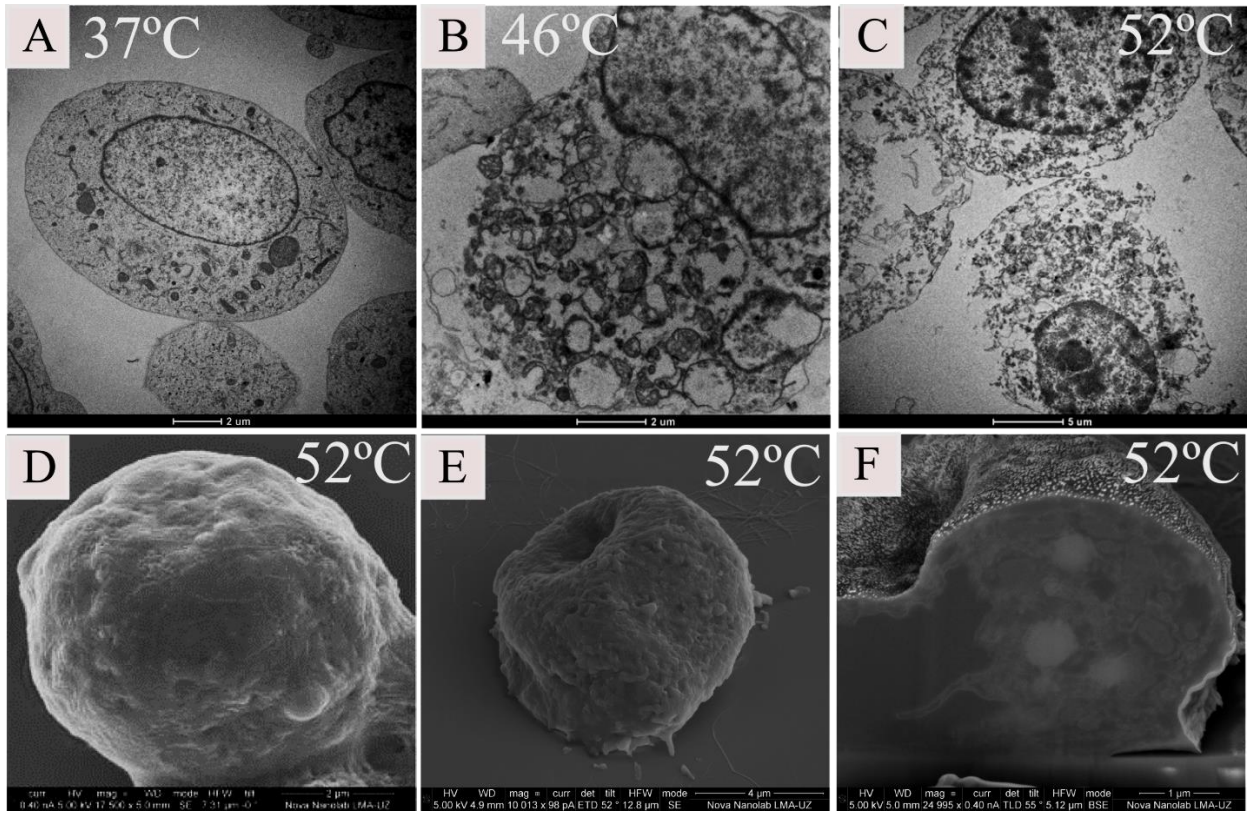

**Figure 5.** Electron micrographs of SH-SY5Y cells after EHT for 30 minutes: A) TEM image of control cell with typical cytological structure; B and C) TEM images of cells after EHT at 46°C (B) and 52°C (C). Note the intracellular vacuolation and the discontinuity of the cytoplasmic membrane related to membrane permeabilization. D) SEM micrograph showing the surface of a single cell after EHT at 52°C. Large membrane invaginations were observed (E), and cell cross-sectional images obtained with SEM-FIB confirmed vacuolation and cytoplasmic damage (F).

*Magnetic hyperthermia (MHT).* Neuroblastoma cells exposed to MHT were also morphologically characterized by TEM and dual-beam FIB-SEM (Figure 6). The control cells exposed to AMF (570 kHz and 23.9 kA/m) for 30 minutes (Figure 6A) presented no detectable morphological changes. These micrographs confirmed that the exposure of cells to AMF in the absence of PEI-MNPs does not produce significant cell damage under our experimental conditions. An analysis of the subcellular organization of these cells revealed well-recognizable organelles in the cytosol, with an endoplasmic reticulum and mitochondria that were easily identifiable, as well as an intact cell membrane and nuclear envelope. Similarly, the cells incubated with PEI-MNPs also exhibited a healthy appearance and displayed visible MNPs grouped inside small membranes and free in the cytosol. In contrast, the cells exposed to MHT at 46°C (Figure 6B and 6C) displayed abnormal membranous stacking dispersed in the cytosol accompanied by cell rounding and a surface membrane with irregular blebbing shapes, which is typical of apoptosis. Interestingly, these features were accompanied by a relatively well-preserved nucleus and an intact cell membrane in most instances. The observed morphologies of the intracellular organelles and cytoplasmic structures were consistent with previous reports.[17,21] Overall, these morphological features suggest cell death by apoptosis.[22] However, some cells also presented several crater-like membrane pores (Figure 6D and 6E) with a diameter of approximately one micron. The absence of PEI-MNPs attached to the cell surface in the majority of the SEM images recorded after MHT led us to hypothesize that these pores could have been caused by particles that were attached to the cell surface during the MHT treatment. We believe that at a temperature of 52°C, the presence of MNPs on the cell surface could have induced a surface heating process that enhanced membrane fluidly and exerted a generalized intracellular membranous wrecking effect. Our analysis of STEM images (SI, Figure S4) supported these results.

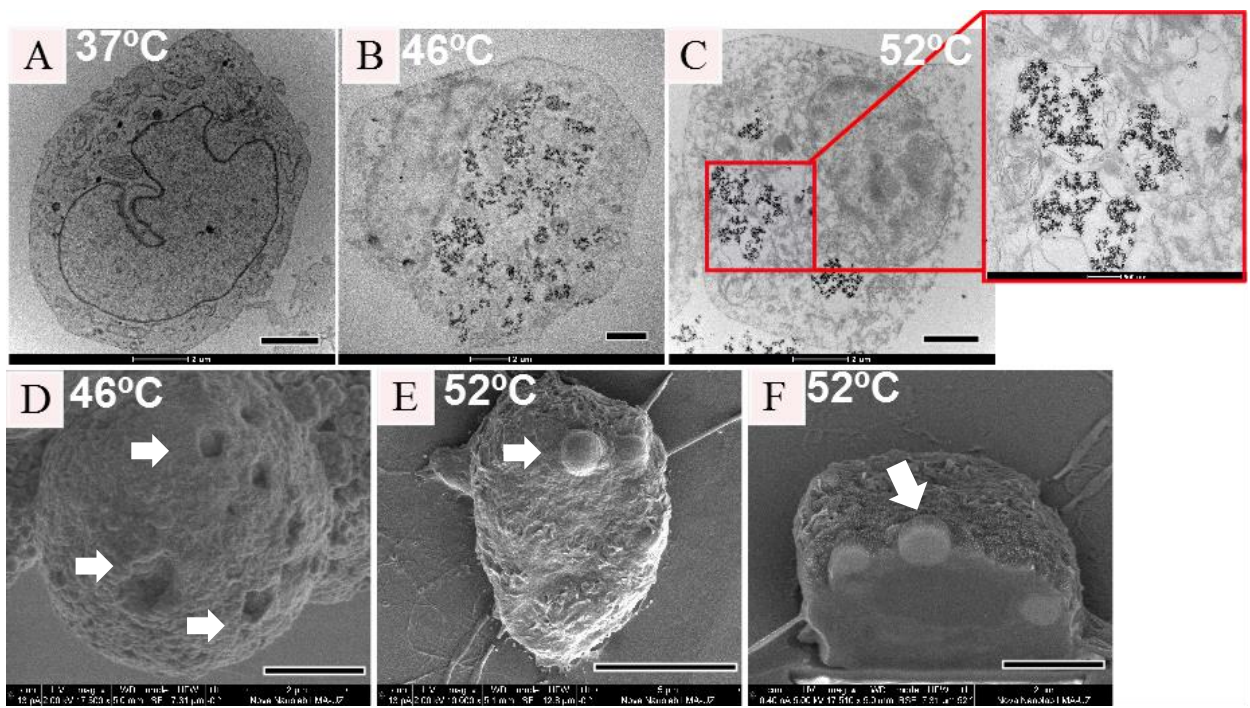

**Figure 6.** TEM and FIB-SEM micrographs of SH-SY5Y cells subjected to MHT for 30 minutes. A) Control cell (AMF, no particles). After MHT at 46°C (B) and 52°C (C), the PEI-MNPs (100 μg/mL) were still easily identifiable within the cytoplasm inside membranous structures. Dual-beam FIB-SEM images after MHT at 46°C (D) show drastic changes in cellular roughness and numerous holes (arrows) and apoptotic blebs on the cytoplasmic membrane (E) and the cytoplasm (F). The scale bars are 2 μm, except in (F) (5 μm).

As in the case of the EHT experiments, a negligible fraction of PEI-MNP clusters were found attached to the cell membranes after MHT. The cellular surface displays typical round blebs, suggesting cell death by apoptosis (Figures 6D-6F, white arrows). These blebs exhibited no Fe signal, as determined through a local EDX spectra analysis (not shown). Most of the population of cells subjected to MHT at 52°C lost their typical round shape and display the typical blebbing of apoptotic cells.

To obtain a complementary overall picture of the intracellular damage induced by MHT, a sample of SH-SY5Y cells treated with MHT at 52°C was dehydrated, contrasted and embedded in epoxy resin to perform a three-dimensional reconstruction from 25-nm-thick cross-sections produced by FIB-SEM. The cellular damage identified from these images and the 3D reconstruction (SI, Figure S5 and video file) extends across the complete cytoplasmatic compartment, and consistent with the above-described results, the distribution of PEI-MNPs after MHT showed that large fractions of these particles were expulsed to the extracellular medium during hyperthermia. The obtained cell ultrastructure showed apparent cytoplasmic cavities and vacuoles surrounded by a damaged cytoplasmic membrane.

It is clear that both heating treatments (EHT and MHT) produced significant thermal damage. However, the cytotoxic effects observed in the MHT-treated neuroblastoma cells were more severe at lower temperatures (T < 42°C). After EHT hyperthermia at these temperatures, cell damage was negligible. Because PEI-MNPs interact strongly with the membrane,[23] we believe that membrane-attached clusters could have exerted local heat-damaging effects in the membrane, producing the membrane pores observed after MHT and triggering necrosis. The ostensibly undisrupted nuclear membrane observed after both treatments might be due to the resistant nuclear membrane cytoskeleton, which assembles through the crosslinking of some intermediate nanometric filaments known as nuclear laminas. These protein networks further support and might preserve the lipid bilayer of the nuclear membrane from disorganizing after the heat shock treatments.

Previous studies have illustrated the confinement of MNPs into endo-lysosomes by TEM[24] or fluorescence confocal microscopy,[25] demonstrating the influence of this confinement on the cell death mechanism experienced by the cultured cells subjected to AMF. Similarly, Asin *et al.*

showed that cell death can be triggered during MHT treatment by the release of countless proteases from the destructed lysosomes into the surrounding cell space.[24] In addition, Creixell *et al.* demonstrated that the action of AMF activates the lysosomal death pathways in cancer cells overexpressing the EGFR.[26] These results suggest that MHT treatments can produce at least two different -and complementary- cell death mechanisms resulting from the actual temperature effect in combination with the effect of (endo)lysosomal destruction. Additionally, Di Corato *et al.*[27] performed *in vitro* and *in vivo* experiments using double-functional (photodynamic and magnetic) liposomes as therapeutic agents and found striking and very promising results regarding complete solid-tumor ablation in a rodent model. These liposomes were primarily located in the plasma membrane at the initial stage and were subsequently internalized within endosomal compartments. Both photodynamic and magnetic mechanisms are required for complete tumor ablation, although only the sub-lethal target temperatures (approximately 40°C) selected for this study could be the origin of the difference between the MHT-alone and dual-modality protocols. Our results, although restricted to *in vitro* experiments, showed that a single MHT modality could be sufficient for complete cell death if the target temperature is adapted to the actual hyperthermia range of 42-45°C. Ogden *et al.*[28] reported a comparative study between MHT and EHT cytotoxicity and demonstrated that the magnetic treatments presented greater effectiveness. This result is not in agreement with the results presented by Jordan *et al.*[17] and Chan *et al.*,[29] who detected no differences between the two hyperthermic approaches. Similarly, Wilhelm *et al.*[30] evaluated the cytotoxic effects induced by EHT and MHT heating for 1 hour, and their results demonstrate that local MHT was as effective as global EHT and suggests a future optimization of the MNP design to improve the heating efficacy.

Our interpretation of the results obtained in the present study is in agreement with that reported by Ogden et al.[28]: MHT is more effective than global heating by EHT because MHT treatment for 30 minutes activates the late apoptosis-necrosis death process. In addition, our nano-heaters (PEI-MNPs) effectively destroyed these 'micro-tumor-like' pellets, which are smaller than those previously reported by Wilhelm and co-workers ($2 \times 10^7$ cells).[30,31] Baba et al.[32] reported differences between superparamagnetic and ferromagnetic particles designed to treat MCF-7 human breast cancer cells and found that ferromagnetic materials exposed to 250 kHz and 4 kW exhibited a higher heating efficiency. Unfortunately, the indication of the power available when performing an AMF experiment is not sufficient to determine the actual magnetic field amplitude delivered; thus, a comparison with our results is not straightforward. The cell death percentages observed when the temperature reached 46.4 and 48.5°C were 50 and 95%, respectively,[32] which are similar to the range of temperatures obtained with our EHT experiments. The mortalities obtained with our *in vitro* MHT results, which were obtained with an application time of 30 minutes, are in the same range as those observed at 43°C by Guardia et al.,[21] who found approximately 50% mortality after 1 hour of treatment (110 kHz and 20 kA/m) and the induction of necrotic damage in KB cells. Yoo et al.[33] reported the effect of magnetic hyperthermia treatment on breast cancer cells (MDA-MB-231) incubated with 200 µg/mL MNPs. These cells were irradiated at 500 kHz and 37.4 kA/m at a temperature of 43°C for 60 minutes. The resulting viability was 75%, which is lower than that observed in our study, and their SPA value was 471 W/g, which is higher than the SPA of our PEI-MNPs.

CONCLUSIONS

Through a systematic series of *in vitro* studies, we compared the different effects that EHT and MHT heating protocols have on cell viability in similar 'micro-tumor-like' environments. We obtained solid evidence that MHT requires lower target temperatures (approximately 6°C lower) to produce the same cell death effect as the EHT protocol, demonstrating the MHT protocol's effectiveness and potential to remotely destroy tumor cells. The MNPs selected for these experiments showed good efficiency for heating these small-volume tumor-phantoms by AMF. Compared with EHT, the cytotoxic effects of MHT were more significant and occurred at lower temperatures. Although the morphological effects suggest that similar cell death pathways are induced by EHT and MHT, in the latter case, the rapid permeabilization and destruction of the cell membrane due to heating by the attached MNP clusters could add to the overall mechanisms induced by the temperature treatment. This study demonstrates how nano-heaters can destroy not just cultures of heat-stress-resistant neuroblastoma cells but also solid masses of neuroblastoma cells, which are comparable to micro-tumors and similar to those cells that initiate tumoral growth after metastasis *in vivo*. Taking into account (i) that magnetic particles are reasonably biocompatible and (ii) that this tumor type is typically characterized by a high degree of metastasis, the future use of nano-heaters could be a major advancement in the treatment of this still incurable pathology. The low target temperature required to kill 50% of the cell population using the MHT protocol, $T_0 = 42.1°C$, is a very promising result, indicating the possibility that it could be developed into therapeutic protocols in clinical oncology.

METHODS

**Synthesis of PEI-MNPs.** PEI-MNP colloids were synthetized using a modified oxidative hydrolysis method, i.e., precipitation of an iron salt ($FeSO_4$) in basic media (NaOH) with a mild oxidant.[34] The method was modified to produce an *in situ* polymer coating with 25-kDa polyethylenimine (PEI) during the reaction to obtain PEI-coated MNPs. The process for synthesizing the water-based magnetic colloid has been reported elsewhere.[23,35]

**Cell culture and compacted pellet conditions.** Human neuroblastoma SH-SY5Y cells (ATCC CRL-2266) were cultured in Dulbecco's Modified Eagle's Medium and Ham's F12 (1:1) with 15% fetal bovine serum, 100 IU/mL penicillin, 100 μg/mL streptomycin and 2 mM L-glutamine. The cells were maintained at 37°C in a saturated humidity atmosphere containing 95% air and 5% $CO_2$. To reduce the intrinsic biological variability and standardize the treatment conditions, we developed a neuroblastoma cell phantom that mimics some very basic conditions of a micro-tumor environment with the aim of emulating *in vivo* treatment experiments. Cellular pellets composed of approximately $6-7 \times 10^6$ neuroblastoma cells immersed in 100 μL of protein-rich culture medium were treated inside PCR microcentrifuge tubes.

**Exogenous hyperthermia treatment.** Exogenous hyperthermia was performed using a water bath (Medingen Labortechnik MWB5). Cells ($1 \times 10^6$ cells/well) were seeded in a six-well plate in 2 mL of culture medium and incubated for two days to reach the exponential growth phase. The cells were washed in PBS, detached and resuspended in 100 μL of supplemented DMEM. The samples were exposed to different temperatures (37 to 56°C) by immersion in a water bath for 30 minutes and then cultured for an additional 6 hours. To observe the long-term effects of the hyperthermia treatment, the viabilities of the cells were calculated immediately after the treatment and after re-culture using the trypan blue (TB) exclusion protocol.

**Magnetic hyperthermia treatment.** Magnetic hyperthermia (MHT) experiments were performed using a commercial DM-1 applicator (nB Nanoscale Biomagnetics S.L., Spain) with constant target-temperature programs, which are applicable for different time periods. The set point was reached asymptotically from room temperature, with very small overheating values (< 1°C). The heating rates and set point temperatures were maintained through feedback-based controlled field amplitude modulation (SI, Figure S6). The experiments were performed at 570 kHz, and the magnetic field amplitudes were varied in the range of $3.98 \leq H \leq 23.9$ kA/m to maintain the target temperature. A predefined number of starting cells ($1 \times 10^6$ cells/well) were seeded in a six-well plate in 2 mL of complete culture medium. The next day, the culture medium was replaced with PEI-MNPs (100 µg/mL) dispersed in supplemented DMEM, and the plate was incubated for 24 hours. After this incubation time, the cells were washed in PBS, detached and resuspended in 100 µL of supplemented DMEM. Compacted cell pellets incubated with PEI-MNPs were placed in DM-1 for 30 minutes and were then cultured for 6 hours. The cell viability was determined using the same protocol that was used with the EHT-treated cells.

**Transmission electron microscopy (TEM).** Samples were prepared by seeding cells ($1 \times 10^6$ cells/well) in a six-well plate in 2 mL of culture medium. Twenty-four hours after plating, MNPs at a concentration of 100 µg/mL were added to the culture. After overnight incubation, the cells were detached and exposed to temperatures of 46 and 52°C for 30 min using both the EHT and MHT heating approaches. Cell samples were fixed with 2% glutaraldehyde for 2 hours at 4°C, treated with 2.5% potassium ferrocyanate and 1% osmium tetroxide, dehydrated and embedded in epoxy resin. Ultrathin sections of slices (70 nm) were cut and observed in a FEI Tecnai F30 microscope operated at an accelerating voltage of 300 kV. The images in STEM-HAADF mode were obtained using a HAADF (high-angle annular dark field) detector for the STEM mode and an EDX (energy-dispersive X-ray) spectrometer.

**Dual-beam (FIB-SEM) analysis.** The intracellular distribution of MNPs in conditioned SH-SY5Y neuroblastoma samples was analyzed using a dual-beam FIB-SEM (Nova 200 NanoLab, FEI Company). SEM images were recorded at 5 and 30 kV using a FEG column. A combined Ga-based 30-kV (10-pA) ion beam was used to cross-section single cells. Dehydrated cells on coverslips were coated with platinum for FIB-SEM imaging. The cell sections embedded in epoxy resin used for the TEM observations were also investigated using a dual-beam Helios 650 instrument from FEI Company, which was operated in the "slide and view" mode using 30 kV (0.43 nA) for the ion beam, and BSE images were acquired at 2 kV and 0.2 nA using the TLD detector in the immersion mode. 3D reconstruction images were processed using Amira software.

**Confocal microscopy imaging.** PEI-MNPs were functionalized with Alexa 488, which is suited to the 488-nm laser line. First, PEI-MNPs were suspended in 1 mL of 0.1 M sodium carbonate-bicarbonate buffer at pH 9.5 to ensure deprotonation of the amine groups of the PEI polymer coating.[36] Alexa 488 solution (1 mg·mL$^{-1}$ DMSO: 20 µL) was then added to the suspension of PEI-MNPs, and the reaction mixture was covered with foil and rotated at room temperature for 3 hours. The final Alexa 488-functionalized PEI-MNPs were washed with deionized water until no further fluorescence was observed in the supernatant. These particles are henceforth denoted as *f*-PEI-MNPs. Following incubation with the particles, cells were exposed to both types of heating treatments (EHT and MHT) and then immediately fixed in 4% paraformaldehyde. Staining of cell actin was performed using phalloidin-tetramethylrhodamine B isothiocyanate (Sigma-Aldrich), and DNA (nucleus and chromosomes) was stained using Hoechst dye (bisbenzimide) (Sigma-Aldrich). Confocal microscopy images were obtained using a Nikon A1R confocal microscope and processed with NIS-Elements Advanced Research software. All confocal cell images were pseudocolored. Ethidium bromide staining was performed for necrosis determination.

**Quantitative/qualitative flow cytometry determination of apoptosis and necrosis.** An average of 10,000 cells per condition were analyzed by flow cytometry using a FACS Aria Cytometer. The data were analyzed using FACS Diva software (Becton Dickinson, NJ, USA). A quantitative analysis of cell death by apoptosis and necrosis was performed using a commercial kit (Annexin V FITC kit, Immunostep, Spain) that can discriminate live cells from those in the early or late phases of apoptosis or necrosis. Three cell populations were detected: (i) unstained cells, representing viable cells, (ii) Annexin V (+) / PI (-) cells, representing early apoptotic cells, and (iii) Annexin V (+) / PI (+) cells, representing late apoptotic or necrotic cells. Early necrosis was determined through membrane permeabilization assays complemented with confocal and electron microscopy examinations.

ASSOCIATED CONTENT

**Supporting Information Available:**

Figure S1 presents curves of the cell viability as a function of the target temperature at different times (15 minutes and 6 hours) after the exogenous heating (EHT) hyperthermia experiments, and the difference between both curves shows a peak at T = 52°C.

Figure S2 presents curves of the cell viability as a function of the target temperature at different times (15 minutes and 6 hours) after the magnetic hyperthermia (MHT) experiments, and the difference between both curves shows a peak at T = 46°C.

Figure S3 presents the dot plots of the cell viability data as obtained from the FACS experiments, in the original format, for control SH-SY5Y cells and also after EHT and MHT heated at the same $T_0$= 42.5ºC target temperature.

Figure S4 presents STEM images of SH-SY5Y cells after MHT at two different target temperatures.

Figure S5 shows snapshots of the 3D reconstruction from cross-sectional images obtained from dual-beam SEM-FIB analysis and illustrates the spatial distribution of the MNPs within and surrounding the cells after 30 minutes of MHT at 52°C.

Figure S6 shows the typical profiles of the temperature and magnetic field control as a function of time, employed to maintain the target temperature constant during the SPA experiments.

The attached movie provides an animated 3D reconstruction of a PEI-MNP-loaded SH-SY5Y cell after MHT for 30 minutes at T = 52°C.

This material is available free of charge via the Internet at http:

**Data Availability**

There are no restrictions whatsoever on the data reported in this work.

ASSOCIATED CONTENT

**Present Addresses**

† Currently at nB Nanoscale Biomagnetics S.L., Zaragoza, Spain.

**Conflict of Interest Disclosure**

G.F.G. and M.R.I., along with the University of Zaragoza, have filled patents related to the technology and intellectual property related to the magnetic field applicators reported herein. The other authors declare that they do not have any affiliations that would lead to conflicts of interest.

**Funding Sources**

This work was supported by the Spanish Ministerio de Economia y Competitividad (MINECO) through project MAT2013-42551.

ACKNOWLEDGMENTS

We are indebted to Dr. E. Campos for his help with the TEM imaging experiments. The technical support provided by LMA-INA and SAI-UZ is also acknowledged.

REFERENCES

(1) Leopold, L. Thyro-Endocrine Hyperthermia. *C. R. Seances Soc. Biol. Ses Fil.* **1919**, *82*, 344-346.

(2) Abe, M.; Hiraoka, M.; Takahashi, M.; Egawa, S.; Matsuda, C.; Onoyama, Y.; Morita, K.; Kakehi, M.; Sugahara, T. Multiinstitutional Studies on Hyperthermia Using an 8-Mhz Radiofrequency Capacitive Heating Device (Thermotron Rf-8) in Combination with Radiation for Cancer-Therapy. *Cancer* **1986**, *58*, 1589-1595.

(3) Mahmoudi, K.; Hadjipanayis, C. G. The Application of Magnetic Nanoparticles for the Treatment of Brain Tumors. *Front. Chem. (Lausanne, Switz.)* **2014**, *2*, 109.

(4) Maier-Hauff, K.; Rothe, R.; Scholz, R.; Gneveckow, U.; Wust, P.; Thiesen, B.; Feussner, A.; von Deimling, A.; Waldoefner, N.; Felix, R.; Jordan, A. Intracranial Thermotherapy Using Magnetic Nanoparticles Combined with External Beam Radiotherapy: Results of a Feasibility Study on Patients with Glioblastoma Multiforme. *J. Neuro-Oncol.* **2007**, *81*, 53-60.

(5) Maier-Hauff, K.; Ulrich, F.; Nestler, D.; Niehoff, H.; Wust, P.; Thiesen, B.; Orawa, H.; Budach, V.; Jordan, A. Efficacy and Safety of Intratumoral Thermotherapy Using Magnetic Iron-Oxide Nanoparticles Combined with External Beam Radiotherapy on Patients with Recurrent Glioblastoma Multiforme. *J. Neuro-Oncol.* **2011**, *103*, 317-324.

(6) Goya, G. F.; Asín, L.; Ibarra, M. R. Cell Death Induced by Ac Magnetic Fields and Magnetic Nanoparticles: Current State and Perspectives. *Int. J. Hyperthermia* **2013**, *29*, 810-818.

(7) Misra, S. K.; Ghoshal, G.; Gartia, M. R.; Wu, Z.; De, A. K.; Ye, M.; Bromfield, C. R.; Williams, E. M.; Singh, K.; Tangella, K. V.; Rund, L.; Schulten, K.; Schook, L. B.; Ray, P. S.;

Burdette, E. C.; Pan, D. Trimodal Therapy: Combining Hyperthermia with Repurposed Bexarotene and Ultrasound for Treating Liver Cancer. *ACS Nano* **2015**, *9*, 10695-10718.

(8) Hildebrandt, B.; Wust, P.; Ahlers, O.; Dieing, A.; Sreenivasa, G.; Kerner, T.; Felix, R.; Riess, H. The Cellular and Molecular Basis of Hyperthermia. *Critical Reviews in Oncology Hematology* **2002**, *43*, 33-56.

(9) Mukherjee, A.; Castanares, M.; Hedayati, M.; Wabler, M.; Trock, B.; Kulkarni, P.; Rodriguez, R.; Getzenberg, R. H.; DeWeese, T. L.; Ivkov, R. Monitoring Nanoparticle-Mediated Cellular Hyperthermia with a High-Sensitivity Biosensor. *Nanomedicine (London, U. K.)* **2014**, *9*, 2729-2743.

(10) Tartaj, P.; Morales, M. D.; Veintemillas-Verdaguer, S.; Gonzalez-Carreno, T.; Serna, C. J. The Preparation of Magnetic Nanoparticles for Applications in Biomedicine. *J. Phys. D: Appl. Phys.* **2003**, *36*, R182-R197.

(11) Mathew, D. S.; Juang, R.-S. An Overview of the Structure and Magnetism of Spinel Ferrite Nanoparticles and Their Synthesis in Microemulsions. *Chem. Eng. J. (Amsterdam, Neth.)* **2007**, *129*, 51-65.

(12) Gazeau, F.; Levy, M.; Wilhelm, C. Optimizing Magnetic Nanoparticle Design for Nanothermotherapy. *Nanomedicine (London, U. K.)* **2008**, *3*, 831-844.

(13) Ito, A.; Shinkai, M.; Honda, H.; Yoshikawa, K.; Saga, S.; Wakabayashi, T.; Yoshida, J.; Kobayashi, T. Heat Shock Protein 70 Expression Induces Antitumor Immunity During Intracellular Hyperthermia Using Magnetite Nanoparticles. *Cancer Immunol. Immunother.* **2003**, *52*, 80-88.

(14) Gilany, K.; Van Elzen, R.; Mous, K.; Coen, E.; Van Dongen, W.; Vandamme, S.; Gevaert, K.; Timmerman, E.; Vandekerckhove, J.; Dewilde, S.; Van Ostade, X.; Moens, L. The Proteome of the Human Neuroblastoma Cell Line Sh-Sy5y: An Enlarged Proteome. *Biochim. Biophys. Acta, Proteins Proteomics* **2008**, *1784*, 983-985.

(15) Xu, G.; Stevens, S. M.; Kobiessy, F.; Brown, H.; McClung, S.; Gold, M. S.; Borchelt, D. R. Identification of Proteins Sensitive to Thermal Stress in Human Neuroblastoma and Glioma Cell Lines. *PLoS One* **2012**, *7*, e49021.

(16) Cheng, L.; Smith, D. J.; Anderson, R. L.; Nagley, P. Human Neuroblastoma Sh-Sy5y Cells Show Increased Resistance to Hyperthermic Stress after Differentiation, Associated with Elevated Levels of Hsp72. *Int. J. Hyperthermia* **2011**, *27*, 415-426.

(17) Jordan, A.; Scholz, R.; Wust, P.; Schirra, H.; Schiestel, T.; Schmidt, H.; Felix, R. Endocytosis of Dextran and Silan-Coated Magnetite Nanoparticles and the Effect of Intracellular Hyperthermia on Human Mammary Carcinoma Cells in Vitro. *J. Magn. Magn. Mater.* **1999**, *194*, 185-196.

(18) Feng, Y.; Oden, J. T.; Rylander, M. N. A Two-State Cell Damage Model under Hyperthermic Conditions: Theory and in Vitro Experiments. *J. Biomech. Eng.* **2008**, *130*, 041016-041016.

(19) Hedayati, M.; Thomas, O.; Abubaker-Sharif, B.; Zhou, H.; Cornejo, C.; Zhang, Y.; Wabler, M.; Mihalic, J.; Gruettner, C.; Westphal, F.; Geyh, A.; Deweese, T. l.; Ivkov, R. The Effect of Cell Cluster Size on Intracellular Nanoparticle-Mediated Hyperthermia: Is It Possible to Treat Microscopic Tumors? *Nanomedicine (London, U. K.)* **2013**, *8*, 29-41.

(20) Nikoletopoulou, V.; Markaki, M.; Palikaras, K.; Tavernarakis, N. Crosstalk between Apoptosis, Necrosis and Autophagy. *Biochim. Biophys. Acta, Mol. Cell Res.* **2013**, *1833*, 3448-3459.

(21) Guardia, P.; Di Corato, R.; Lartigue, L.; Wilhelm, C.; Espinosa, A.; Garcia-Hernandez, M.; Gazeau, F.; Manna, L.; Pellegrino, T. Water-Soluble Iron Oxide Nanocubes with High Values of Specific Absorption Rate for Cancer Cell Hyperthermia Treatment. *ACS Nano* **2012**, *6*, 3080-3091.

(22) Burattini, S.; Falcieri, E. Analysis of Cell Death by Electron Microscopy. In *Necrosis*, McCall, K.; Klein, C., Eds.; Humana Press: New York, 2013; Vol. 1004, pp 77-89.

(23) Calatayud, M. P.; Sanz, B.; Raffa, V.; Riggio, C.; Ibarra, M. R.; Goya, G. F. The Effect of Surface Charge of Functionalized Fe3o4 Nanoparticles on Protein Adsorption and Cell Uptake. *Biomaterials* **2014**, *35*, 6389-6399.

(24) Asin, L.; Goya, G. F.; Tres, A.; Ibarra, M. R. Induced Cell Toxicity Originates Dendritic Cell Death Following Magnetic Hyperthermia Treatment. *Cell Death Dis.* **2013**, *4*, e596.

(25) Treuel, L.; Malissek, M.; Gebauer, J. S.; Zellner, R. The Influence of Surface Composition of Nanoparticles on Their Interactions with Serum Albumin. *ChemPhysChem* **2010**, *11*, 3093-3099.

(26) Creixell, M.; Bohorquez, A. C.; Torres-Lugo, M.; Rinaldi, C. Egfr-Targeted Magnetic Nanoparticle Heaters Kill Cancer Cells without a Perceptible Temperature Rise. *ACS Nano* **2011**, *5*, 7124-7129.

(27) Di Corato, R.; Bealle, G.; Kolosnjaj-Tabi, J.; Espinosa, A.; Clement, O.; Silva, A. K. A.; Menager, C.; Wilhelm, C. Combining Magnetic Hyperthermia and Photodynamic Therapy for Tumor Ablation with Photoresponsive Magnetic Liposomes. *ACS Nano* **2015**, *9*, 2904-2916.

(28) Ogden, J.; Tate, J.; Strawbridge, R.; Ivkov, R.; Hoopes, P. Comparison of Iron Oxide Nanoparticle and Waterbath Hyperthermia Cytotoxicity. In *Energy-based Treatment of Tissue and Assessment V*, Proceedings of SPIE 7181, San Jose, CA, Jan 24, 2009; Ryan, T. P., Ed. International Society for Optics and Photonics, 2009; p 71810K.

(29) Chan, D. C. F.; Kirpotin, D. B.; Bunn Jr., P. A. Synthesis and Evaluation of Colloidal Magnetic Iron Oxides for the Site-Specific Radiofrequency-Induced Hyperthermia of Cancer. *J. Magn. Magn. Mater.* **1993**, *122*, 374-378.

(30) Wilhelm, C.; Fortin, J.-P.; Gazeau, F. Tumour Cell Toxicity of Intracellular Hyperthermia Mediated by Magnetic Nanoparticles. *J. Nanosci. Nanotechnol.* **2007**, *7*, 2933-2937.

(31) Di Corato, R.; Espinosa, A.; Lartigue, L.; Tharaud, M.; Chat, S.; Pellegrino, T.; Menager, C.; Gazeau, F.; Wilhelm, C. Magnetic Hyperthermia Efficiency in the Cellular Environment for Different Nanoparticle Designs. *Biomaterials* **2014**, *35*, 6400-6411.

(32) Baba, D.; Seiko, Y.; Nakanishi, T.; Zhang, H.; Arakaki, A.; Matsunaga, T.; Osaka, T. Effect of Magnetite Nanoparticles on Living Rate of Mcf-7 Human Breast Cancer Cells. *Colloids Surf., B* **2012**, *95*, 254-257.

(33) Yoo, D.; Jeong, H.; Preihs, C.; Choi, J.-s.; Shin, T.-H.; Sessler, J. L.; Cheon, J. Double-Effector Nanoparticles: A Synergistic Approach to Apoptotic Hyperthermia. *Angew. Chem., Int. Ed. Engl.* **2012**, *51*, 12482-12485.

(34) Verges, M. A.; Costo, R.; Roca, A. G.; Marco, J. F.; Goya, G. F.; Serna, C. J.; Morales, M. P. Uniform and Water Stable Magnetite Nanoparticles with Diameters around the Monodomain-Multidomain Limit. *J. Phys. D: Appl. Phys.* **2008**, *41*, 134003.

(35) Sanz, B.; Calatayud, M. P.; Cassinelli, N.; Ibarra, M. R.; Goya, G. F. Long-Term Stability and Reproducibility of Magnetic Colloids Are Key Issues for Steady Values of Specific Power Absorption over Time. *Eur. J. Inorg. Chem.* **2015**, *2015*, 4524-4531.

(36) Fang, C.; Veiseh, O.; Kievit, F.; Bhattarai, N.; Wang, F.; Stephen, Z.; Li, C.; Lee, D.; Ellenbogen, R. G.; Zhang, M. Functionalization of Iron Oxide Magnetic Nanoparticles with Targeting Ligands: Their Physicochemical Properties and in Vivo Behavior. *Nanomedicine (London, U. K.)* **2010**, *5*, 1357-1369.

Table of Contents graphic

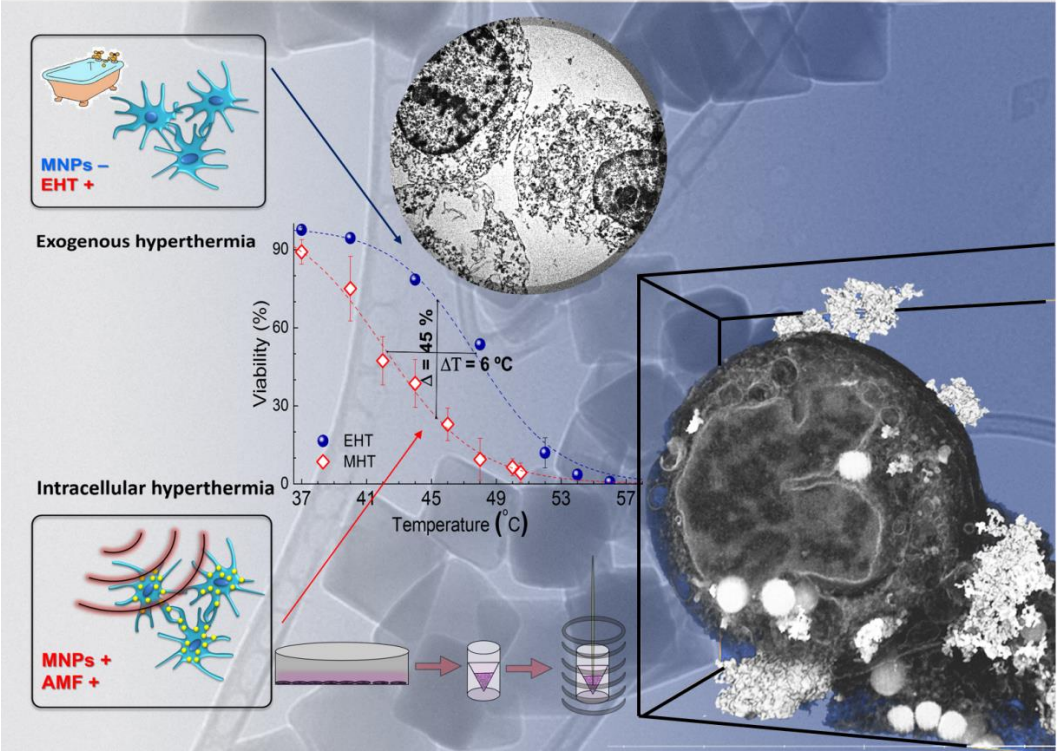